# Simulated annealing with time-varying strain in a dipole-coupled array of magnetostrictive nanomagnets

Md Ahsanul Abeed and Supriyo Bandyopadhyay

Department of Electrical and Computer Engineering

Virginia Commonwealth University

Richmond, VA 23284, USA

In a two-dimensional arrangement of closely spaced elliptical nanomagnets with in-plane magnetic anisotropy, whose major axes are aligned along columns and minor axes along rows, dipole coupling will make the magnetic ordering "ferromagnetic" along the columns and "anti-ferromagnetic" along the rows. Noise and other perturbations can drive the system out of this ground state configuration and pin it in a metastable state where the magnetization orientations will not follow this pattern. Internal energy barriers, sufficiently larger than the thermal energy kT, will prevent the system from leaving the metastable state and decaying spontaneously to the ground state. These barriers can be temporarily eroded by globally straining the nanomagnets with time-varying strain if the nanomagnets are magnetostrictive, which will allow the system to return to ground state after strain is removed. This is a hardware emulation of simulated annealing in an interacting many body system. Here, we demonstrate this function experimentally.

Consider a system of elliptical nanomagnets with in-plane magnetic anisotropy arranged as in Fig. 1 on a substrate. Because of the shape anisotropy, each nanomagnet will have an easy axis along the major axis, which will make its magnetization orient along one of two opposite directions along the major axis. Dipole interaction between the nanomagnets will result in the configuration shown in Fig. 1, where all nanomagnets along a column are magnetized in the same direction along the major axis, but alternating columns have opposite (anti-parallel) magnetizations. This is the ground state configuration.

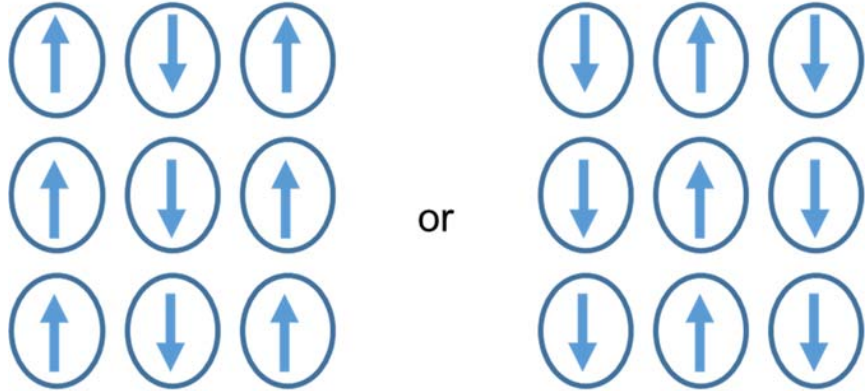

Fig. 1: Magnetization orientations in an array of dipole coupled elliptical nanomagnets

In order to verify that Fig. 1 is indeed the ground state configuration, we calculated the potential energy for a system of $3 \times 3$ array of elliptical cobalt nanomagnets of major axis dimension 350 nm, minor axis dimension 320 nm and thickness 12 nm, using the micromagnetic simulator MuMax3 which takes into account the demagnetizing field due to shape anisotropy, exchange interaction within a nanomagnet and dipole interaction between the nanomagnets. Since there are 9 nanomagnets in the $3 \times 3$ array, each with 2 possible orientations of the magnetization, there are $2^9 = 512$ possible combinations corresponding to 512 possible magnetic configurations. The energies of these combinations are plotted in Fig. 2. Clearly, there are two (degenerate) minimum energy configurations and they conform exactly to the two shown in Fig. 1.

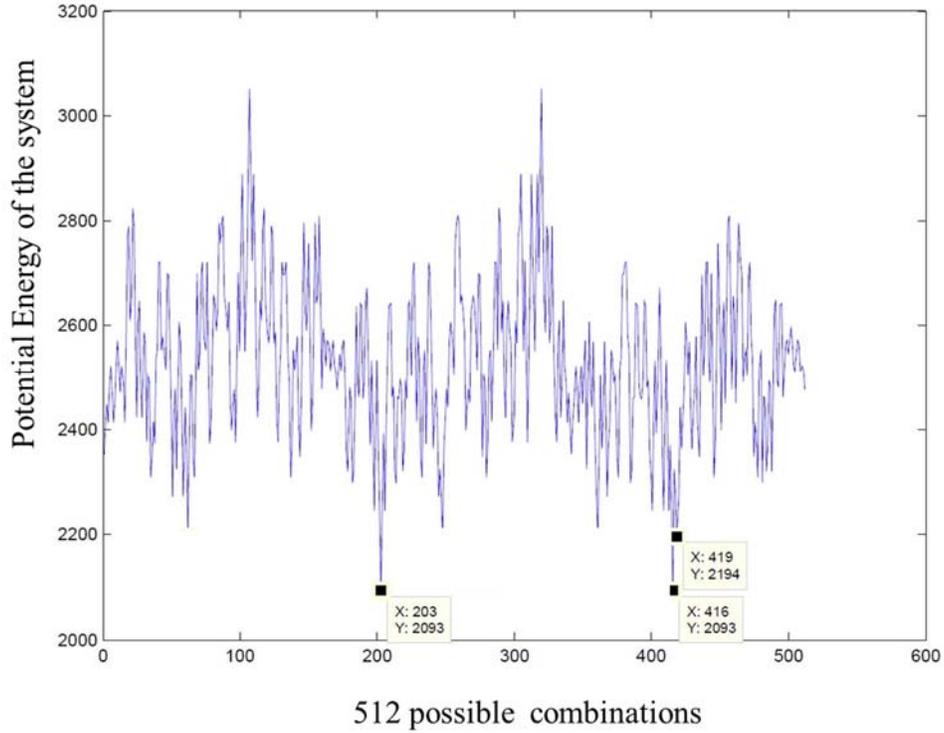

Fig. 2: Potential energies of various possible combinations of magnetic ordering in a 3 × 3 array of elliptical cobalt nanomagnets

In Fig. 3, we show the magnetic ordering of the 3 × 3 array, with the micromagnetic distribution within each nanomagnet, for one of the two ground state configurations.

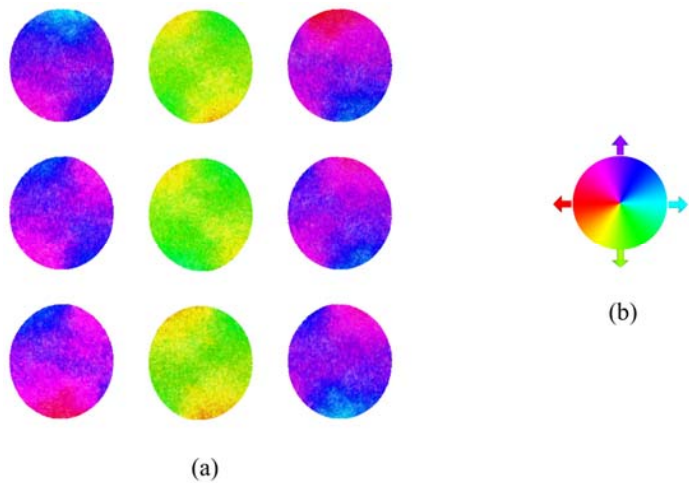

(a)

(b)

Fig. 3: (a) The computed magnetic ordering in a 3 × 3 array of elliptical cobalt nanomagnets. (b) Color wheel for the micromagnetic distribution within any nanomagnet (the color denotes the local magnetization direction within any nanomagnet)

Suppose now that the magnetizations are perturbed by an external agent (noise, stray magnetic field, etc.) which drives the system out of the ground state and destroys the ground state magnetic ordering in Fig. 1 or 3. The system may not be able to return spontaneously to the ground state if there are metastable states (caused by e.g. small variations in magnet shape or thickness, and/or structural defects such as material voids) and the system gets stuck in one of them. Internal energy barriers that separate the metastable state from the ground state will prevent the system to transition to the ground state. In this case, supplying energy from outside, or eroding the intervening energy barriers with an external agent, will allow the system to transition to the ground state. We can view this as a hardware emulation of simulated annealing [1] since the process allows the system to unpin itself and migrate from the metastable to the ground state.

In a magnetostrictive nanomagnet system, *strain* can be the external agent that triggers the simulated annealing action. To understand why strain has this effect, consider the cartoon in Fig. 4(a) where we show the (arbitrary) potential profile inside a nanomagnet (potential energy versus magnetization orientation). We have assumed that the nanomagnet has imperfections such as edge roughness, thickness variations, material defects, etc. that cause the potential profile to have one or more metastable states that would be absent in an ideal (perfect) nanomagnet. There is a global energy minimum and some local energy minima, which are arbitrary in this figure because they are caused by random defects. Note that because of dipole coupling, one orientation along the major axis ($\theta = 180^0$) is slightly preferred over the other ($\theta = 0^0$).

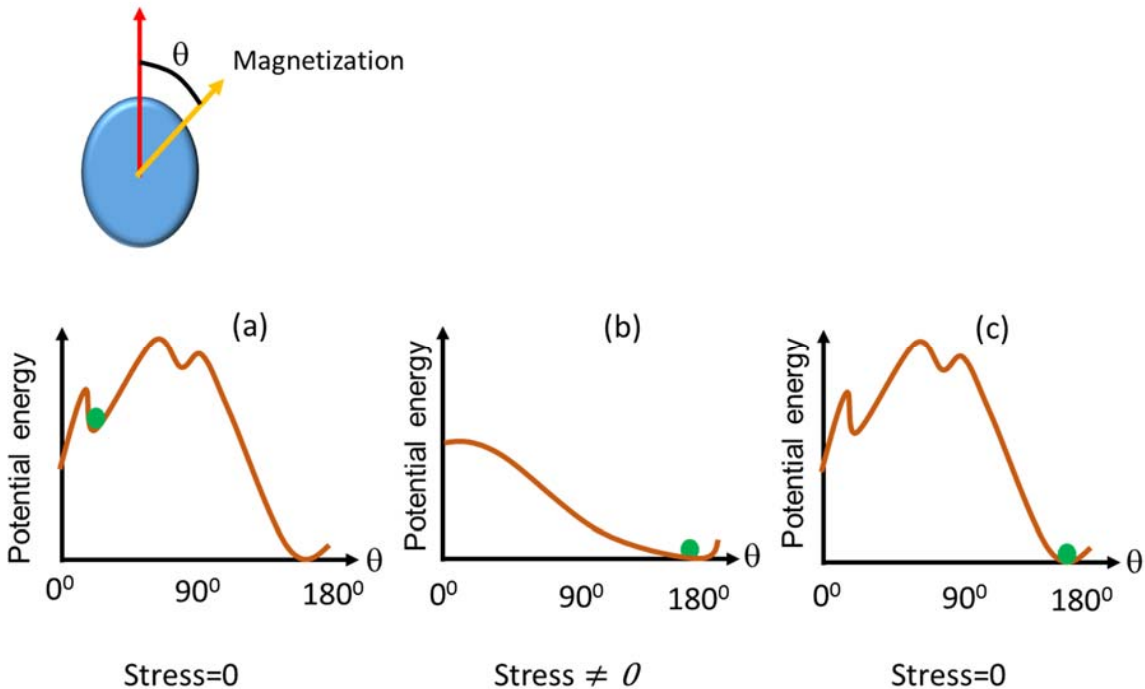

Fig. 4: Potential energy profile in a nanomagnet where $\theta$ denotes the magnetization orientation. (a) Owing to perturbation, the magnetization is stuck in a metastable state as denoted by the ball, (b) potential energy profile in the presence of stress where the barriers are removed and the magnetization can relax to the ground state, and (c) potential energy profile after removal of stress where the magnetization remains in the ground state.

If we perturb the magnetizations of one or more nanomagnets and drive the array out of the global energy minimum and pin it into a local minimum, the system will get stuck in the metastable state and cannot decay to the ground state because of the potential barriers separating the metastable and ground states as shown in Fig. 4(a). However, when we apply compressive or tensile strain, the energy landscape is altered and may look like the one in Fig. 4(b) as long as the product of the magnetostriction coefficient and strain is a negative quantity. Strain alters the potential landscape. Once that happens, the energy barrier(s) between the metastable and ground states is(are) eroded and the system can relax to the ground state configuration as shown in Fig. 4(c).

To test this model, we fabricated cobalt nanomagnets on a piezoelectric LiNbO₃ substrate using electron beam patterning of a resist, electron beam evaporation of cobalt on to the patterned substrate and lift-off. An atomic force micrograph of a 3 × 3 array is shown in Fig. 5. The dimensions conform to the ones used for the simulation: major axis = 350 nm, minor axis = 320 nm and thickness = 12 nm.

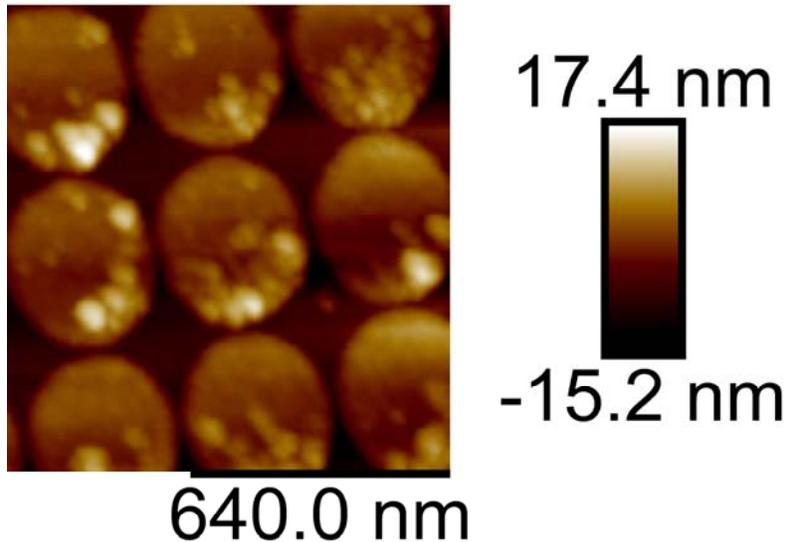

Fig. 5: Atomic force micrographs of elliptical cobalt nanomagnets fabricated on a LiNbO₃ substrate.

We then determined the magnetic ordering with magnetic force microscopy (MFM). The MFM image of the 3 × 3 array is shown in Fig. 6(a). We clearly see the ordering computed in Fig. 3, where the nanomagnets along a column are all magnetized in the same direction and alternating columns have opposite directions of magnetization. This image is obtained with a low-moment tip in order to carry out non-invasive imaging. Next, we intentionally perturb the magnetization in the array with a high-moment tip and we show the MFM image of the resulting configuration after the perturbation in Fig. 6(b). Clearly the ground state ordering has been destroyed and the system has not spontaneously returned to the ground state (it is stuck in a metastable state). We then launch a surface acoustic wave (SAW) in the substrate, which periodically exerts tensile and compressive strain on the nanomagnets and rotates their magnetization via the Villari effect [2-20]. The SAW is launched by delineating side electrodes on the piezoelectric substrate and applying a sinusoidal voltage of 24 V and frequency is 3.57 MHz to one electrode. After the SAW excitation is terminated, we image the nanomagnets again, and find that the system has returned to the ground state. The SAW temporarily eroded the potential barriers that impeded transition from the metastable state to the ground state and allowed the system to relax to the ground state. This is an emulation of simulated annealing. Here, the periodic strain (SAW) acted as the simulated annealing agent.

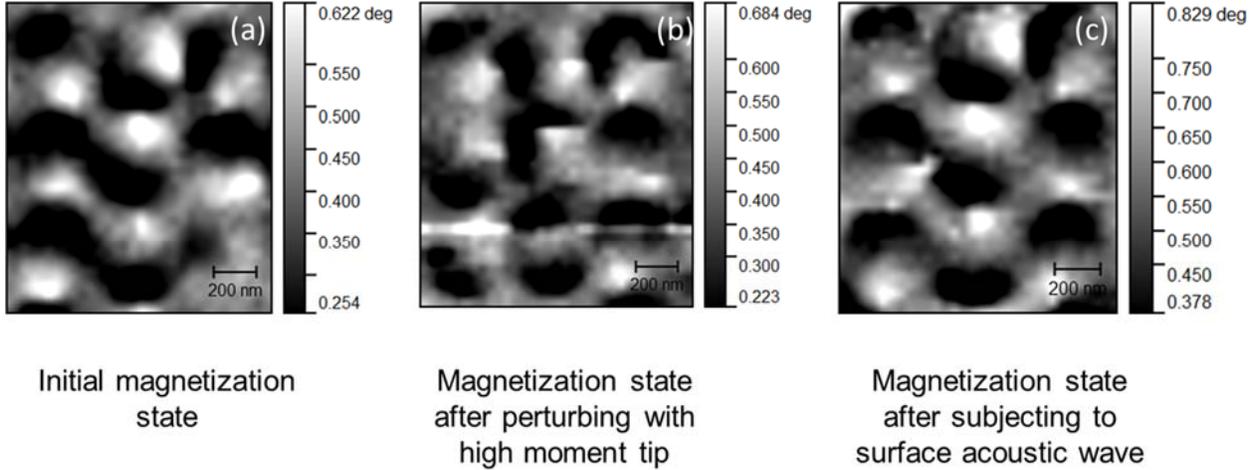

Initial magnetization state

Magnetization state after perturbing with high moment tip

Magnetization state after subjecting to surface acoustic wave

Fig. 6: Magnetic force microscopy images showing (a) the initial magnetization state, (b) the state after perturbation with a high moment tip and (c) the magnetization state after passage of the surface acoustic wave.

One interesting question is whether we can perturb the system enough that it goes from one ground state configuration to the other after removal of the perturbation. There are two issues with this. First, perturbing with a high moment tip probably does not provide enough perturbation for this to happen. Second, the two ground states are degenerate only in an ideal system, but in a non-ideal system, they may have slightly different energies, resulting in a monostable system rather than a bistable system. In our case, the system always returned to the original state after removal of perturbation and did not transition to the other ground state.

In conclusion, we have shown that in an interacting system of magnetostrictive nanomagnets (interacting via dipole coupling), time-varying strain associated with a surface acoustic wave acts as an agent of simulated annealing. This is an example of hardware based simulated annealing. It may have applications in hardware accelerators to solve combinatorial optimization problems.

**Acknowledgement**: This work is partially supported by the US National Science Foundation under grants ECCS-1609303 and CCF-1815303. The authors are indebted to Mr. Dhritiman Bhattacharya for advice with MFM.

# References

1. *Quantum Annealing and Related Optimization Methods*, Eds. A. Das and B. K. Chakrabarti (Springer-Verlag, Berlin Heidelberg, 2005).

2. V. Sampath, N. D'Souza, D. Bhattacharya, G. M. Atkinson, S. Bandyopadhyay and J. Atulasimha, *Nano Lett.*, **16**, 5681-5687 (2016).

3. V. Sampath, N. D'Souza, G. M. Atkinson, S. Bandyopadhyay and J. Atulasimha, *Appl. Phys. Lett.*, **109**, 102403 (2016).

4. K. Roy, S. Bandyopadhyay and J. Atulasimha, *arXiv:1012.0819,* (2012).

5. A. K. Biswas, S. Bandyopadhyay and J. Atulasimha, *Appl. Phys. Lett.*, **103**, 232401 (2013).

6. S. Davis, A. Baruth and S. Adenwalla, *Appl. Phys. Lett.*, **97**, 232507 (2010).

7. W. Li, B. Buford, A. Jander and P. Dhagat, *IEEE Trans. Magn.*, **50**, 2285018 (2014).

8. W. Li, B. Buford, A. Jander and P. Dhagat, *J. Appl. Phys.,* **115**, 17E307 (2014).

9. O. Kovalenko, T. Pezeril and V. V. Temnov, *Phys. Rev. Lett.*, **110**, 266602 (2013).

10. M. Foerster, F. Macià, N. Statuto, S. Finizio, A. Hernàndez-Mínguez, S. Lendínez, P. V. Santos, J. Fontcuberta, J. M. Manel Hernàndez, M. Kläui and L. Aballe, *Nat. Commun.*, **8**, 407 (2017).

11. M. Bombeck, A. S. Salasyuk, B. A. Glavin, A. V. Scherbakov, C. Brüggemann, D. R. Yakovlev, V. F. Sapega, X. Liu, J. K. Furdyna, A. V. Akimov and M. Bayer, *Phys. Rev. B*, **85**, 195324 (2012).

12. A. V. Scherbakov, A. S. Salasyuk, A. V. Akimov, X. Liu, M. Bombeck, C. Brüggemann, D. R. Yakovlev, V. F. Sapega, J. K. Furdyna and M. Bayer, *Phys. Rev. Lett.,* **105**, 117204 (2010).

13. L. Thevenard, I. S. Camara, S. Majrab, M. Bernard, P. Rovillain, A. Lemaître, C. Gourdon and J.-Y. Duquesne, *Phys. Rev. B,* **93**, 134430 (2016).

14. M. Weiler, L. Dreher, C. Heeg, H. Huebl, R. Gross, M. S. Brandt and S. T. B. Goennenwein, *Phys. Rev. Lett.*, **106**, 117601 (2011).

15. J. Janusonis, C. L. Chang, P. H. M. van Loosdrecht and R. I. Tobey, *Appl. Phys. Lett.*, **106**, 181601 (2015).

16. U. Singh and S. Adenwalla, *Nanotechnology,* **26**, 255707 (2015).

17. L. Thevenard, J-Y. Duquesne, E. Peronne, H. J. von Bardeleben, H. Jaffres, S. Ruttala, J.-M. George, A. Lemaître and C. Gourdon, *Phys. Rev. B*, **87**, 144402 (2013).

18. E. M. Chudnovsky and R. Jaafar, *Phys. Rev. Appl.*, **5**, 031002 (2016).

19. J. Tejada, E. M. Chudnovsky, R. Zarzuela, N. Statuto, J. Calvo-de la Rosa, P. V. Santos and A. Hernández-Mínguez, *Europhys. Lett.*, **118**, 37005 (2017).

20. S. Mondal, M. A. Abeed, K. Dutta, A. De, S. Sahoo, A. Barman and S. Bandyopadhyay, *ACS Appl. Mater, Interfaces*, **10**, 43970 (2018).